\documentclass[twocolumn]{el-author}

\newcommand{\ua}{\uparrow}
\newcommand{\nc}{\newcommand}
\nc{\da}{\downarrow} \nc{\hc}{\hat{c}} \nc{\hS}{\hat{S}}
\nc{\bra}{\langle} \nc{\ket}{\rangle} \nc{\eq}{equation (\ref}
\nc{\h}{\hat} \nc{\hT}{\h{T}}\nc{\be}{\begin{eqnarray}}
\nc{\ee}{\end{eqnarray}}\nc{\rd}{\textrm{d}}\nc{\e}{eqnarray}\nc{\hR}{\hat{R}}\nc{\Tr}{\mathrm{Tr}}
\nc{\tS}{\tilde{S}}\nc{\tr}{\mathrm{tr}}\nc{\8}{\infty}\nc{\lgs}{\bra\ua,\phi|}\nc{\rgs}{|\ua,\phi\ket}
\nc{\hU}{\hat{U}}\nc{\lfs}{\bra\phi|}\nc{\rfs}{|\phi\ket}\nc{\hZ}{\hat{Z}}\nc{\hd}{\hat{d}}\nc{\mD}{\mathcal{D}}
\nc{\bd}{\bar{d}}\nc{\bc}{\bar{c}}\nc{\mc}{\mathcal}\nc{\ea}{eqnarray}\nc{\mG}{\mathcal{G}}\nc{\bce}{\begin{center}}
\nc{\ece}{\end{center}}
\date{12th December 2011}

\begin{document}

\title{Compressive sensing based differential channel feedback for massive MIMO}

\author{Wenqian Shen, Linglong Dai, Yi Shi, Xudong Zhu, and Zhaocheng Wang, \textit{IET Fellow}}

\abstract{Massive multiple-input multiple-output (MIMO) is becoming a key technology for future 5G wireless communications. Channel feedback for massive MIMO is challenging due to the substantially increased dimension of MIMO channel matrix.
In this letter, we propose a compressive sensing (CS) based differential channel feedback scheme to reduce the feedback overhead.
Specifically, the temporal correlation of time-varying channels is exploited to generate the differential channel impulse response (CIR) between two CIRs in neighboring time slots, which enjoys a much stronger sparsity than the original sparse CIRs.
Thus, the base station can recover the differential CIR from the highly compressed differential CIR under the framework of CS theory.
Simulations show that the proposed scheme reduces the feedback overhead by about 20\% compared with the direct CS-based scheme.}

\maketitle

\section{Introduction}
Massive multiple-input multiple-output (MIMO) is emerging as a key technology for future 5G wireless communications.
For massive MIMO systems, accurate channel state information at transmitter (CSIT) is essential to improve both the spectrum and energy efficiency \cite{scaling_up_mimo}.
One approach to acquire CSIT is to adopt time division duplexing (TDD) mode by exploiting the channel reciprocity.
However, as frequency division duplexing (FDD) outperforms TDD in delay-sensitive and traffic-symmetric applications, FDD still dominates current cellular networks without channel reciprocity \cite{Shim}, where effective channel feedback is crucial to enable the expected system performance of massive MIMO.
As the dimension of the MIMO channel matrix becomes very large in massive MIMO systems, while the dedicated resource for channel feedback is limited in the uplink, channel feedback becomes a challenging problem.
Conventional channel feedback methods based on codebook are not suitable for massive MIMO systems,
since the size of the codebook has to be extensively expanded to guarantee an acceptable CSIT accuracy \cite{protocal}.
For non-codebook based channel feedback, which usuallly requires large overhead for channel feedback, compressive sensing (CS) has been applied to reduce the feedback overhead in massive MIMO systems \cite{protocal}-\cite{CIR_CS}.
Some CS-based channel feedback schemes \cite{protocal}, \cite{CSIT} exploited the spatial correlations among antennas to compress the MIMO channel matrix for overhead reduction, but MIMO channels are not always spatially correlated in practice, especially when distributed MIMO is considered \cite{scaling_up_mimo}.
In \cite{CIR_CS}, the channel impulse response (CIR) is directly compressed by exploiting the time-domain channel sparsity to reduce the feedback overhead, but the overhead is still high, especially when the CIR is not sufficiently sparse.

In this letter, we propose a CS-based differential feedback scheme to reduce the feedback overhead by exploiting the temporal correlation of MIMO channels. Specifically, as the temporal channel correlation exists in both centralized and distributed MIMO systems, we propose to generate the differential CIR by differentiating the CIRs in two adjacent time slots, which enjoys a much stronger sparsity than the original CIRs. We then propose to compress the differential CIR with lower compression ratio, which equivalently requires much lower feedback overhead.
Simulation results show that to achieve the same normalized mean square error (NMSE) performance,
the proposed scheme can reduce the feedback overhead by about 20\% than the direct CS-based channel feedback.

\section{Temporal correlation of time-varying MIMO channels}
In the $t$-th time slot, the CIR between the $n$-th transmit antenna at the base station (BS) and the single-antenna user in massive MIMO systems can be denoted as $\mathbf{h}_n^{(t)}=[h_n^{(t)}(0),h_n^{(t)}(1),\cdots,h_n^{(t)}(L-1)]$ for $1\leq n \leq N$, where $N$ is the number of transmit antenna and $L$ is the maximal channel delay spread.
The CIR is often considered to be sparse (i.e., there are $K$ non-zero elements in $\mathbf{h}_n^{(t)}$, $1\leq K\ll L$) since it usually contains a few significant paths in broadband communications \cite{Guiguan}.
The sparse CIR series $\{\mathbf{h}_n^{(t)}\}_{t=1}^T$ in $T$ consecutive time slots exhibits temporal correlation even when the MIMO channels are fast time-varying \cite{Dai13}.
We model such temporal correlation through a variation of the CIR's support (i.e., the positions of non-zero elements) and an evolution of the amplitudes of non-zero elements.
To be specific, the time-varying sparse CIR can be characterized by a support vector $\mathbf{p}_n^{(t)}$ and an amplitude vector $\mathbf{a}_n^{(t)}$ as follows
\begin{align}
  \mathbf{h}_n^{(t)}=\mathbf{p}_n^{(t)}\circ\mathbf{a}_n^{(t)},
\end{align}
where $p_n^{(t)}(l)\in\{0,1\}$, $a_n^{(t)}(l)\in\mathcal{C}$ and $\circ$ denotes the Hadamard product.
To model the variations of the CIR's support $\mathbf{p}_n^{(t)}$ over time, the $l$-th elements of $T$ support vectors $\{p_n^{(t)}(l)\}_{t=1}^T$ in consecutive $T$ time slots can be modeled as a first-order Markov process \cite{Dynamic_CS},
which can be fully characterized by two transition probabilities $p_{10}\triangleq \text{Pr}\{p_n^{(t+1)}(l)=1|p_n^{(t)}(l)=0\}$ and $p_{01}\triangleq \text{Pr}\{p_n^{(t+1)}(l)=0|p_n^{(t)}(l)=1\}$,
and a distribution $\mu_n^{(1)}\triangleq \text{Pr}\{p_n^{(1)}(l)=1\}$ in the initial time slot $t=1$.
For the steady-state Markov process, where $\text{Pr}\{p_n^{(t)}(l)=1\}=\mu, \forall t,n$, only two parameters $p_{01}$ and $\mu$ are enough to characterize such process since $p_{10}=\mu p_{01}/(1-\mu)$.
Then, to model the evolution of the CIR's amplitudes over time, the amplitude vector $\mathbf{a}_n^{(t)}$ can be modeled by the first-order autoregressive model as follows \cite{Dynamic_CS}
 \begin{align}
  \mathbf{a}_n^{(t)}=\rho\mathbf{a}_n^{(t-1)}+\sqrt{1-\rho^2}\mathbf{w}^{(t)},
\end{align}
where $\rho =J_0(2\pi f_d\tau)$ is the correlation coefficient given by the zero-order Bessel function of the first kind with $f_d$ being the maximal Doppler frequency and $\tau$ being the time slot duration,
the parameter $\mathbf{w}^{(t)}$ denotes a noise vector independent of $\mathbf{a}_n^{(t-1)}$, and its entities are independent and identically distributed (i.i.d.) zero-mean complex Gaussian variables following the distribution $\mathcal{CN}(0,\sigma_{\omega}^2)$.
\section{Compressive sensing based differential channel feedback}
As shown in Fig. 1, unlike the direct CS-based channel feedback scheme, which directly compresses the sparse CIR with a sensing matrix, the proposed CS-based differential channel feedback scheme further exploits the temporal correlation of MIMO channels to generate the differential CIR which enjoys a much stronger sparsity than the original CIRs.
The differential CIR is then compressed by the sensing matrix, which is known to both the BS and the users,
and then CS algorithms can be used to produce a precise recovery of the differential CIR after channel feedback.
In order to avoid feedback error propagation, we further propose that users execute initialization after a certain (either fixed or adaptive) number of time slots, whereby the original CIR is compressed with a relatively higher compression ratio to ensure a precise initial CIR recovery.
More specifically, the procedure of the proposed scheme can be decomposed into the following three steps.
\begin{figure}[h]
\vspace{-3mm}
\centering{\includegraphics[width=85mm]{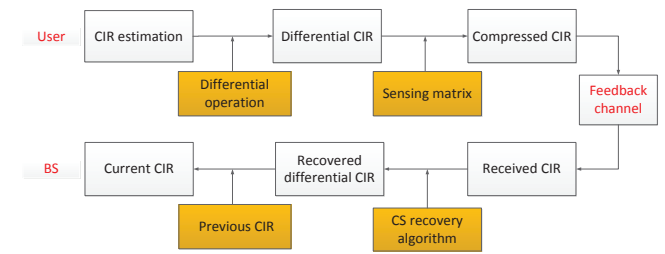}}
\vspace{-2mm}
\caption{Compressive sensing based differential channel feedback.}
\vspace{-5mm}
\end{figure}
\section{1. Differential operation}
Unlike conventional feedback schemes which directly feedback the estimated channel in the current time slot, the proposed scheme aims to create a more sparse representation of the channel by computing the difference between two estimated CIRs $\mathbf{h}_n^{(t-1)}$ and $\mathbf{h}_n^{(t)}$ in the previous $(t-1)$-th and the current $t$-th time slots, respectively. Mathematically, the differential CIR $\Delta \mathbf{h}_n^{(t)}$ can be calculated as
 \begin{align}
  \Delta \mathbf{h}_n^{(t)}&=\mathbf{h}_n^{(t)}-\mathbf{h}_n^{(t-1)}\nonumber\\ &=\mathbf{p}_n^{(t)}\circ(\mathbf{a}_n^{(t)}-\mathbf{a}_n^{(t-1)})+(\mathbf{p}_n^{(t)}-\mathbf{p}_n^{(t-1)})\circ\mathbf{a}_n^{(t-1)}\\ &=\mathbf{p}_n^{(t)}\circ[\sqrt{1\!-\!\rho^2}\mathbf{w}\!-\!(1\!-\!\rho)\mathbf{a}_n^{(t-1)}]+(\mathbf{p}_n^{(t)}\!-\!\mathbf{p}_n^{(t-1)})\circ\mathbf{a}_n^{(t-1)}\nonumber.
\end{align}
The first item on the last line of $(3)$ is very small due to the smooth evolution of the CIR's amplitudes, e.g., $\rho\approx1$ \cite{Dynamic_CS}.
The number of non-zero elements of the second item is also small (e.g., not larger than $K$) since the sudden disappearance and appearance of a significant path do not occur often.
Fig. 2 presents the snapshot of the previous CIR, the current CIR, and the differential CIR of the channels described by (1) and (3) with the following parameters: $L=200$, $p_{01}=0.05$, $\mu=0.1$, $fd=10$Hz, $\tau=1$ms, and $\sigma_w=1$.
The initial amplitude $\mathbf{a}_n^{(1)}$ is a random vector, the entries of which are i.i.d. complex Gaussian variables with zero mean and unit variance, and the initial support $\mathbf{p}_n^{(1)}$ is a random vector, the entries of which are i.i.d. Bernoulli variables.
We can observe from Fig. 2 that the differential CIR enjoys a much stronger sparsity than the original CIR due to the temporal correlation of channels, which indicates that it can be compressed more to reduce the feedback overhead.
\begin{figure}[h]
\vspace{-7mm}
\centering{\includegraphics[width=73mm]{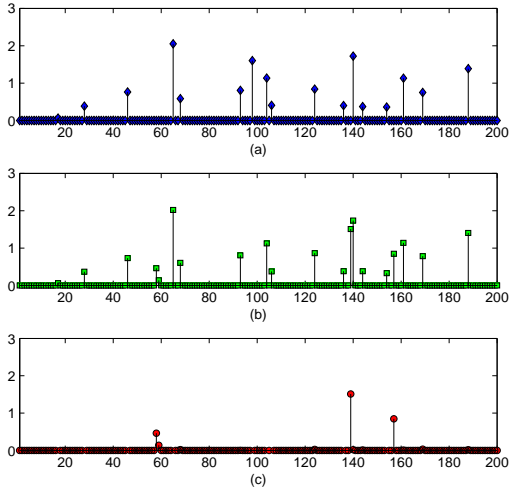}}
\vspace{-5mm}
\caption{A snapshot of the previous CIR, the current CIR, and the differential CIR: (a) the previous CIR; (b) the current CIR; (c) the differential CIR.}
\vspace{-3mm}
\end{figure}
\section{2. CIR compression and recovery}
CS theory shows that a sparse signal $\Delta\mathbf{h}_n^{(t)}$ can be compressed by a sensing matrix $\mathbf{\Phi}\in\mathcal{C}^{M\times L}$ with $M\ll L$ into a low-dimensional measurement vector $\mathbf{y}=\mathbf{\Phi}\Delta\mathbf{h}_n^{(t)}$ ($\mathbf{y}$ will be fed back to the BS).
If $\mathbf{\Phi}$ satisfies the restricted isometry property (RIP) \cite{Qi14},
the sparse signal $\Delta\mathbf{h}_n^{(t)}$ can be reliably recovered by the BS from the noisy measurements $\mathbf{y}=\mathbf{\Phi}\Delta\mathbf{h}_n^{(t)}+\mathbf{n},$ where $\mathbf{n}$ denotes the noise of feedback channel,
the entities of which are i.i.d. complex Gaussian variables following the distribution $\mathcal{CN}(0,\sigma_n^2)$.
In this letter, we set $\mathbf{\Phi}$ as the commonly used random Gaussian matrix \cite{SubspacePursuit}.
As for the CS signal recovery algorithm, we adopt the widely used algorithm called subspace pursuit (SP) due to its low complexity and robustness to noise \cite{SubspacePursuit}.
\section{3. Error propagation avoidance}
Precise recovery of the CIR in the initial time slot is of great importance since the channel feedback error in early stage will propagate in the following CIR recovery process.
Additionally, an unexpected recovery error of the differential CIR will also impair the subsequent CIR recovery process.
To avoid such error propagation, the proposed feedback scheme will be reinitiated every $P$ time slots.
Although such initialization will cause relatively large feedback overhead to ensure a precise recovery of the initial CIR, it will be verified numerically in the next part that the highly compressed differential CIR will save much more feedback overhead in the subsequent time slots.
Thus, the total feedback overhead of the proposed scheme is still lower than the direct CS-based feedback scheme.
\section{Simulation results}
Simulations have been conducted to validate the NMSE performance of the proposed CS-based differential channel feedback scheme.
We consider a massive MIMO system with the same parameters used for the generation of Fig. 2 with $N=32$ antennas. The initial amplitude vector $\mathbf{a}_n^{(1)}$ and support vector $\mathbf{p}_n^{(1)}$ for $1\leq n\leq N$ are independent of each other.
We denote the channel compression ratio (i.e., relative feedback overhead) as $\eta=M/L$, and two cases are considered: $\eta=25\%$ and $\eta=45\%$.
For the proposed scheme, $\eta$ is denoted as the average compression ratio over $P$ feedback slots.
The compression ratios in the initial time slot and subsequent time slots are $45\%$ and $15\%$ for the first case, and $65\%$ and $35\%$ for the second case. Thus, the average compression ratio of the proposed scheme is $25\%$ for the first case and $45\%$ for the second case when $P=3$ is considered as an example (Note that P can be adaptive to the channel condition).
As shown in Fig. 3, the proposed scheme outperforms the direct CS-based scheme when the same compression ratio is considered in both cases.
In the first case, the direct CS-based scheme fails to work since the feedback overhead is insufficient, while the proposed scheme performs well.
In the second case, the proposed scheme achieves a 4 dB SNR gain compared with the direct CS-based scheme.
Additionally, the proposed scheme with $\eta=25\%$ and the direct CS-based scheme with $\eta=45\%$ have very similar NMSE performance, which indicates that the proposed scheme can significantly reduce the feedback overhead from 45\% to 25\% to ensure a reliable channel feedback.
\begin{figure}[h]
\vspace{-3mm}
\centering{\includegraphics[width=90mm]{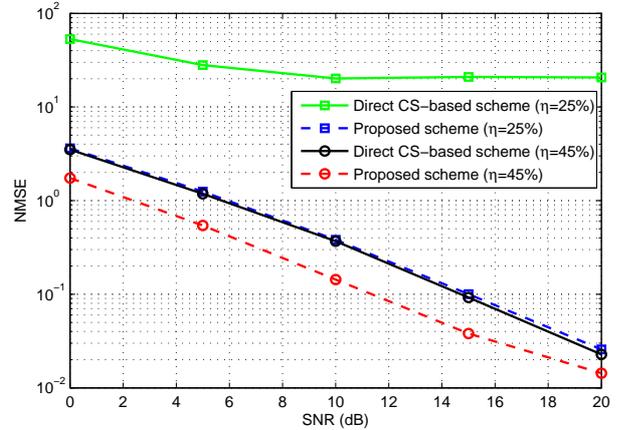}}
\vspace{-5mm}
\caption{NMSE performance comparison between the direct CS-based feedback scheme and the proposed feedback scheme.}
\vspace{-3mm}
\end{figure}
\section{Conclusion}
This letter investigates the challenging problem of channel feedback in massive MIMO systems. It is found that by exploiting the temporal correlation of time-varying channels, the generated differential CIR between CIRs in neighboring time slots enjoys a much stronger sparsity than original CIRs. We have shown that the proposed CS-based differential channel feedback scheme can achieve better NMSE performance than the direct CS-based channel feedback scheme, and reduce the feedback overhead by about 20\% for massive MIMO systems.

\vskip5pt

\noindent W. Shen, L. Dai (Corresponding author), X. Zhu, and Z. Wang (\textit{Tsinghua National Laboratory for Information Science and Technology, Department of Electronic Engineering, Tsinghua University, Beijing 100084, China})

\noindent Yi Shi (\textit{Huawei Technologies, Beijing 100095, China})
\vskip3pt

\noindent E-mail: daill@tsinghua.edu.cn

\end{document}